**Title**

# To mask or not to mask? Investigating the impact of accounting for spatial frequency distributions and susceptibility sources on QSM quality

**Authors and affiliations**

Anders Dyhr Sandgaard[1], Noam Shemesh[2], Sune Nørhøj Jespersen[1,3], Valerij G. Kiselev[4]

[1]Center for Functionally Integrative Neuroscience, Department of Clinical Medicine, Aarhus University, Denmark

[2]Champalimaud Research, Champalimaud Centre for the Unknown, Lisbon, Portugal,

[3]Department of Physics and Astronomy, Aarhus University, Denmark

[4]Division of Medical Physics, Department of Radiology, University Medical Center Freiburg, Freiburg, Germany

**Corresponding Author**

Valerij G. Kiselev

Mail: valerij.kiselev@uniklinik-freiburg.de

# Abstract

**Purpose**

Estimating magnetic susceptibility using MRI depends on inverting a forward relationship between the susceptibility and measured Larmor frequency. However, an often-overlooked constraint in susceptibility fitting is that the Larmor frequency is only measured inside the sample, and after successful background field removal, susceptibility sources should only reside inside the same sample. Here we test the impact of accounting for these constraints in susceptibility fitting.

**Methods**

Two different digital brain phantoms with scalar susceptibility were examined. We used the MEDI phantom, a simple phantom with no background fields, to examine the effect of the imposed constraints for various levels of SNR. Next, we considered the QSM reconstruction challenge 2.0 phantom with and without background fields. We estimated the parameter accuracy of openly-available QSM algorithms by comparing fitting results to the ground truth. Next, we implemented the mentioned constraints and compared to the standard approach.

**Results**

Including the spatial distribution of frequencies and susceptibility sources decreased the root-mean-square-error compared to standard QSM on both brain phantoms when background fields were absent. When background field removal was unsuccessful, as is presumably the case in most in vivo conditions, it is better to allow sources outside the brain.

**Conclusion**

Informing QSM algorithms about the location of susceptibility sources and where Larmor frequency was measured improves susceptibility fitting for realistic SNR levels and efficient background field removal. However, the latter remains the bottleneck of the algorithm. Allowing for external sources regularizes unsuccessful background field removal and is currently the best strategy in vivo.

## Introduction

Magnetic susceptibility $\chi(r)$ can vary greatly across different tissue types[1], making it a highly desirable contrast mechanism reflecting the local chemical composition in biological tissues. A method for mapping $\chi(r)$ with MRI is dubbed quantitative susceptibility mapping[2–8] (QSM), which aims to determine a voxel-specific scalar magnetic susceptibility with promising results[9–11]. In QSM, the examined tissue is effectively assumed to be isotropic with slowly varying magnetic susceptibility. This assumption allows the Larmor frequency shift, $\Omega(r)$, to be written as a convolution of the susceptibility $\chi(r)$ with a Lorentz-corrected dipole kernel[12]. Extensions to this simplified model incorporating heterogenous tissue microstructure have also been proposed[13–15].

The inverse problem of estimating $\chi(r)$ from the measured $\Omega(r)$ has proven to be very difficult. In particular, the inversion is singular as the dipole kernel contains a zero-valued cone. Besides the ill-posed nature of the inversion, and irrespective of the assumed model for tissue complexity, an additional challenge arises from the limited volume in which the Larmor frequency is measured[16], which we refer to as the sample. Moreover, due to the non-local effects of susceptibility, the sample in fact does not include all sources of the induced magnetic field within it: for example, magnetization induced in the body affects the Larmor frequency measured in the brain. The effect of such sources is reduced by so-called background field removal techniques[17], but present QSM algorithms[7,8,18–22] utilizing the Fourier domain to solve the inverse problem result in magnetic susceptibility distributed within the whole field of view (FOV), i.e. not limited to the actual sample volume.

Here, we investigate the effect of informing susceptibility fitting algorithms that the Larmor frequency is only measured inside the sample, and constraining susceptibility sources to the volume in which the Larmor frequency has been measured and corrected by the background field removal. Using digital susceptibility brain phantoms, we evaluate the parameter accuracy of openly-available QSM algorithms for varying peak signal SNR and background fields, with and without the proposed constraint.

## Theory

**Forward problem - finding Ω**

We consider the relationship between the induced Larmor frequency shift $\Omega(r)$ in a sample and measured at discrete sampling positions. For simplicity, we assume each discrete location is characterized by a scalar susceptibility $\chi(r)$ as a result of averaging over the microstructure. Hence, the sample resembles an isotropic media with voxel-wise constant $\chi(r)$, as in conventional QSM. The relationship between $\chi(r)$ and $\Omega(r)$ can then be written as a linear matrix-vector equation

$$\vec{\Omega} = \mathbf{A}\vec{\chi}, \text{ (Forward model)}. \quad (1)$$

Here $\vec{\Omega}$ and $\vec{\chi}$ are N × 1 vectors denoting the measured Larmor frequencies and susceptibility sources in vector form, respectively, where N is the number of voxels in the field of view (FOV). The matrix **A** is an N × N symmetric matrix describing the induced frequency shift in the FOV due to the voxel magnetization in the main field. The diagonal of **A** is zero for isotropic liquids, but if the sample exhibits anisotropic magnetic tissue properties, it acquires a non-zero diagonal[14]; we do not consider such media here. $\mathbf{A}\vec{\chi}$ amounts to a linear convolution[2–4,23] of **A** with $\vec{\chi}$, which can conveniently be implemented in Fourier space using the convolution theorem. Notice that Eq. (1) predicts a non-zero frequency in the whole FOV.

**Inverse problem - finding $\chi$**

The aim of QSM is to estimate $\vec{\chi}$ from the measured $\vec{\Omega}$ using the known **A** in Eq. (1). This amounts to inverting (deconvolving with) **A**, which is a well-known ill-posed problem due to the functional form of the dipole field. The list of algorithms with various weights and regularizations to invert Eq. (1) is long, see e.g. refs. [8,19–22,24–26] for a few examples. Prior to fitting (assuming an ideal signal acquisition), $\vec{\Omega}$ must be extracted from the signal phase. Subsequently, the phase must be unwrapped[17,27]. Finally, contributions from external magnetized sources must be accurately removed for Eq. (1) to represent sources residing inside the sample, a task for which numerous different algorithms exist[17].

If contributions from external sources are successfully eliminated, we are left with the following two points (P1 and P2) of consideration:

**P1)** Eq. (1) predicts a non-zero frequency $\vec{\Omega}$ in the whole FOV, while we only measure it inside the sample due to lack of water outside.

We account for the lack of frequency measurements outside the sample by introducing a sample mask **M** (not to be confused with magnetization), an N × N diagonal matrix with $M_{ii} = 1$ if and only if voxel $i$ is inside the sample. Then the information about the actually measured volume is captured by replacing $\mathbf{A} \to \mathbf{MA}$. This substitution is effectively done when minimizing the SNR-weighted least squares.

**P2)** Susceptibility $\vec{\chi}$ should only be non-zero inside the sample upon successful background-field-removal.

We enforce P2 using the same mask **M**, so $\vec{\chi} \rightarrow \mathbf{M}\vec{\chi}$.

Constraints similar to P1 and P2 have previously been used to remove background fields[28], but here we emphasize that such measures must also be taken for internal sources to represent the measured data correctly. Including P1 and P2 into Eq. (1) forms the new inverse problem we wish to solve, and which is the focus of the present study:

$$\min_{\vec{\chi}} \left\| \vec{\Omega} - \mathbf{MAM}\vec{\chi} \right\|_2^2, \text{ (inverse problem)}. \qquad (2)$$

Here $\|\cdot\|_\vartheta$ denotes the $l_\vartheta$-norm ($\vartheta = 2$ in Eq. (2)). Similar to conventional weighted least squares in QSM[8], **M** limits the utility of solving the inverse problem exclusively in Fourier space as it becomes a convolution also in Fourier space. This suggests the use of iterative least squares to estimate $\vec{\chi}$.

## Methods

**Digital brain phantom simulation**

All simulations were performed in Matlab (The MathWorks, Natick, MA, USA). We tested the minimization problem, Eq. (2), on two different brain phantoms of increasing complexity.

*MEDI phantom*

The first phantom is a digital brain phantom with a spatially varying scalar susceptibility $\vec{\chi}_{\text{GT}}$ provided with the MEDI toolbox[18,24,25,29,30] and their MRI signal generator. This produced a complex multi-echo gradient signal $S(t) = \frac{|\chi_{\text{GT}}|}{\max(|\chi_{\text{GT}}|)} exp(-i\Omega t) + (\varepsilon(t) + i\eta(t))$ with independent Gaussian noise $N\left(0, \frac{1}{\text{SNR}^2}\right)$ in the real and imaginary signal channels for each voxel. We investigated a peak signal SNR ranging from 10 to 200, and with no noise (SNR=∞). Using the MEDI toolbox, the Larmor frequency $\Omega$ used for fitting $\chi$ was estimated based on fitting a complex exponential to the signal. No unwrapping and background-field removal was necessary as the corresponding effects were not simulated.

*QSM reconstruction challenge 2.0 (RC2) phantom*

The second digital phantom considered was the (RC2) phantom[31,32] with a downscaled isotropic resolution of 1 mm and signal SNR of 100. The RC2 phantom includes two versions of MRI complex signals: (a) with no external sources (similar to the MEDI phantom) and (b) including realistic background fields from biological sources outside the brain such as air cavities, bone, fat etc. In both cases the signal magnitude does not directly correspond to susceptibility (in contrast to the MEDI phantom). We extracted the Larmor frequency by fitting a complex exponential to the signal with code supplied with the MEDI toolbox. Then we used SEGUE[27] for phase unwrapping and LBV[33] for background field removal (only in case (b), with depth and peel set to 8).

*Effect on true iIn -vivo and ex -vivo data fitting*

Since true real ex vivo or in vivo images do not have a known ground truth, we chose to focus this study on digital brain phantoms. However, we also investigated the effect of including masks on real MRI data using the QSM reconstruction challenge 1.0 (RC1) in-vivo data with 12 sample orientations and acquired ex-vivo mouse brain data at ultra-high fields (16.4T). For RC1 we implemented masks into COSMOS[8] using the LSMR[34] algorithm.

**Fitting algorithms**

To be on par with current standards, we took openly available iterative QSM algorithms and measured their performance in estimating the ground truth scalar susceptibility $\vec{\chi}_{GT}$ for each digital phantom. Next, we implemented the masks into the algorithms according to Eq. (2) and compared the results to calculations without masking. We investigated MEDI[18,24,25,29,30] and two additional $l1$ and $l2$ regularized iterative algorithms[20], which we denote *Fl1* and *Fl2*, respectively. Including the masks described by P1 and P2, we obtain the following minimization algorithms:

$$\textbf{MEDI}: \min_{\vec{\chi}} \|\mathbf{W}_1 \nabla \vec{\chi}\|_1 + \lambda \left\|\mathbf{W}_2 (\vec{\Omega} - \mathbf{AM}\vec{\chi})\right\|_2^2, \text{(Liu et al, 2012)}, \quad (3)$$

$$\textbf{F}l\textbf{1}: \min_{\vec{\chi}} \alpha \|\nabla \vec{\chi}\|_1 + \frac{1}{2} \left\|(\vec{\Omega} - \mathbf{MAM}\vec{\chi})\right\|_2^2, \text{(Bilgic et al 2012)}, \quad (4)$$

$$\textbf{F}l\textbf{2}: \min_{\vec{\chi}} \beta \|\nabla \vec{\chi}\|_2^2 + \frac{1}{2} \left\|(\vec{\Omega} - \mathbf{MAM}\vec{\chi})\right\|_2^2, \text{(Bilgic et al 2012)}. \quad (5)$$

$\mathbf{W}_1$ is a structural weighting matrix derived from the gradient of the signal magnitude, while $\mathbf{W}_2$ is proportional to signal magnitude to compensate for noise variations (see [18,35]). Hence, $\mathbf{W}_2$ already accounts for **P1** in MEDI, while the $\mathbf{AM}\vec{\chi}$ term is the modification added in this study.

**Analysis and optimal parameter values**

Fitting tolerances, maximum iterations etc. were kept as the default settings for all algorithms (we refer to the source code).

We computed the normalized root-mean-squared-error (RMSE) relative to ground truth across the whole brain for all brain phantoms. Before computing the RMSE, the susceptibility fits, and ground truth were demeaned. For the RC2 phantom, the RMSE was also computed in reference to their mean susceptibility from CSF in ventricles. Additionally, RMSE was computed only in (1) WM, GM and thalamus and (2) deep GM nuclei (DGN). The fitting parameters $\lambda, \alpha, \beta$ in Eqs.(3)-(5) were optimized by the RMSE of the whole brain (optimization shown in Figure S1-S2 in supplementary material).

**Results**

*MEDI phantom*

Figure 1 shows the RMSE between the fitted $\vec{\chi}$ and ground truth $\vec{\chi}_{GT}$ for the MEDI phantom, for all considered methods, and for different SNRs. Across all three algorithms (MEDI, F$l1$, F$l2$), adding masks to the fitting algorithms reduced the error. The largest improvement was clearly in F$l1$ and F$l2$, while MEDI improved a few percent. Figure 2 shows the susceptibility fits and difference from ground truth for 3 different SNRs. Here the improvement is clear for Fl1 and Fl2 for all SNRs, while changes in MEDI were most noticeable in the noiseless case.

*RC2 phantom*

Figure 3 shows the RMSE between the $\vec{\chi}$ and $\vec{\chi}_{GT}$ for the RC2 phantom for the cases with or without external sources (w. BF in bottom row/w.o. BF in upper row, respectively), while Figure 4 shows the corresponding susceptibility fits and difference to ground truth. We observed an improvement for all considered methods when including the masks to the algorithm (red rim is RMSE when fitting with masks), but only in the absence of background fields. When external sources were present, the frequency map after background field removal and unwrapping deviated 44% RMSE compared to the ground truth, even though we eroded the brain 8 layers during BFR. This error carried over to the susceptibility fitting. Hence, allowing external sources provided a form of regularization for the imperfect estimation of background fields.

*Effect on in vivo and ex vivo data fitting*

For in vivo images we found a 24% root-mean-squared difference between demeaned fits with and without masks. Here we report the difference between the two fits due to lack of ground truth. A similar difference was found in ex-vivo data. A full description along with susceptibility fits can be found in the supplementary material (Figure S3-S4). Here the large changes in vivo may be indicative of the unsuccessful background field removal which results in great differences, while this appears to less of a problem ex vivo.

## Discussion

**Simulations**

*MEDI phantom*

Using optimized regularization parameters in the simulations, we found improvements in fitting quality by incorporating masks in the fitting algorithms. For the MEDI phantom, the improvement increased for increasing SNR. For MEDI, the RMSE decreased by 5% when SNR was 100, and in the limit of no noise, the RMSE dropped by 78%. For F$l2$ the decrease was 63% and 75%, respectively, and for F$l1$ the decrease was 70% and 93%, respectively. This demonstrates that even the MEDI algorithm, which is tailored for fitting this exact type of phantom based on its regularization, has visible differences from ground truth solely due to not informing about source localization. The large improvement on F$l1$ and F$l2$ stems from adding both P1 and P2, while for MEDI we only added P2.

*QSM 2.0 reconstruction (RC2) phantom*

In general, we found that referencing to CSF produced the lowest RMSE. This makes sense, as the largest sources of error between the susceptibility fit and the ground truth are from veins. When demeaning, the distributions will then not align optimally, in comparison to referencing to CSF where the susceptibility estimate should be close to the ground truth. For that reason, we discuss here the changes in RMSE wrt. referencing to CSF.

For the RC2 phantom without external sources, we increased the parameter accuracy in susceptibility values even after unwrapping. The highest accuracy was achieved using MEDI Eq. (5), with decreases in RMSE around 5% in the whole brain, 10% in GM and WM tissue, and 2% in DGN. This improvement agrees with the MEDI phantom results since we used an SNR of 100. For a noiseless simulation, we expect this improvement to be higher. F$l1$ decreased in RMSE around 16% in the whole brain, 26% in GM and WM tissue, and 31% in DGN. F*l2* decreased in RMSE around 11% in the whole brain, 12% in GM and WM tissue, and 1% in DGN.

For the RC2 phantom with external sources, where background field removal was needed prior to QSM, we generally see that allowing sources outside produces the lowest RMSE. The highest parameter accuracy was again achieved using MEDI Eq. (5). Here, fitting with the mask increased the RMSE around 13% in the whole brain, 13% in GM and WM tissue, and 8% in DG. For F*l1,* fitting with masks decreased the RMSE around 10% in the whole brain, 26% in GM and WM tissue, and 24% in DGN. For F*l1,* fitting with masks decreased the RMSE below 1% in the whole brain, and increased 6% in GM and WM tissue, and 18% in DGN. In all cases, more regularization was needed when fitting with the constraint on allowed susceptibility sources (cf. figure S2 in supplementary material). Hence, even though we found lower RMSE for F*l1* it was at the expense of increased smoothing of susceptibility maps, which may not be favorable.

**To mask or not to mask?**

The absence of signal outside the sample is an unavoidable feature in MRI, and not including this limitation can lead to erroneous susceptibility estimations. The imposed error can be understood by considering the inverse problem, Eq. (2), using the iterative conjugate gradient method[36]. Here the solution $\vec{\chi}_{k+1}$ at iteration $k+1$, depends on all the previous residual vectors $\vec{r}_k = \vec{\Omega} - \mathbf{A}\vec{\chi}_k$. Since $\mathbf{A}\vec{\chi}_k$ generates a frequency in the whole FOV, no matter the constraint imposed on $\vec{\chi}$, and $\vec{\Omega}$ is zero outside the sample, every residual $\vec{r}_k$ obtains an error, as the fitting algorithm will try to find a solution that reproduces the zeroes in $\vec{\Omega}$, which in turn leads to erroneous solutions $\vec{\chi}_{k+1}$. This error can in principle be avoided by introducing a sample mask $\mathbf{M}$, so the residual vectors spanning the solutions become $\vec{r}_k = \vec{\Omega} - \mathbf{M}\mathbf{A}\vec{\chi}_k$. This constraint can also be included, as in MEDI, through an SNR weight $\mathbf{W}_2$ on the fidelity term.

The second constraint on allowed positions of susceptibility sources arose not directly because of unavoidable limitations in how we measure, but rather due to how $\vec{\Omega}$ is processed before fitting. Namely, as external sources to the sample produce large slowly varying frequency contributions inside the sample, such contributions must be removed prior to susceptibility fitting. While many different algorithms exist for this purpose[17], common to them all is that an ROI (e.g. $\mathbf{M}$) must be defined, where frequency contributions from sources outside the ROI is removed. This leaves us with the natural assumption that the only remaining frequency contributions originate from sources within that ROI. If the ROI is $\mathbf{M}$, the residual vectors in our conjugate gradient example should be $\vec{r}_k = \vec{\Omega} - \mathbf{M}\mathbf{A}\mathbf{M}\vec{\chi}_k$, to avoid sources outside $\mathbf{M}$ when estimating $\vec{\chi}_{k+1}$.

However, when background field removal is incomplete, the susceptibility fits $\vec{\chi}_{k+1}$ will incur an error, and additional errors arise due to the constraint itself. This is due to the estimated Larmor frequency including field errors induced by sources outside the sample, and these errors are optimally "regularized" by allowing for sources outside the sample to compensate.

**Recommendations for QSM fitting**

To mask or not to mask thus depends on the type of study. Digital phantoms are highly popular for practicing solving the inverse problem, and reconstruction challenges[32] have already been carried out. For such studies, no outside sources are to be found, and one can therefore achieve a boost in parameter accuracy by constraining positions of susceptibility. We therefore recommend using the proposed constraints for such studies.

For in vivo QSM, background field removal is paramount, but results in non-negligible errors in the frequency map (for the RC2 phantom, the frequency map had a 44% RMSE compared to the ground truth frequency map without external sources, even after heavy erosion). This indicates the background field removal as the central problem of the current QSM algorithms (disregarding anisotropic tissues). Incomplete removal of the background field implies the presence of residual sources outside the brain. Hence, one will achieve the best susceptibility maps by allowing sources outside. Nevertheless, in all cases, we found that it improved parameter accuracy to inform the algorithm about where the frequency was measured, which is evident from the MEDI algorithm.

**Limitations**

We illustrated the level of improvement achievable in representative existing algorithms, and how easy these corrections can be incorporated. The matrix **A** may also contain additional complexity, for example in terms of mesoscopic effects, which introduces diagonal terms to **A** capturing local structural and/or magnetic anisotropy[13,14].

Simulating the effect of masking on digital phantoms, enabled a comparison to a known ground truth and control over the amount of noise in the MRI signal and the background field. However, while testing major confounds, other artifacts may also hamper the performance boost, e.g., eddy current distortions, sample motion etc. Such effects could be included in the phantom and studied as well.

**Conclusion**

We demonstrated an improvement in susceptibility fitting results for digital brain phantoms by incorporating the constraints that the Larmor frequency is only measured inside the sample. When the phantoms did not include brain-external sources, susceptibility fitting was further improved by constraining susceptibility sources to not reside outside the sample.

**Code availability**

MATLAB code is included in supplementary material giving an overview of the changes implemented in used algorithms.

**Acknowledgments**

We thank Shemesh lab members, especially PhD Cristina Chavarrías and MSc Beatriz Cardoso for assisting with MRI experiments, Prof. Mark D Does and Dr. Kevin Harkins from Vanderbilt University for the REMMI pulse sequence, and Prof. Jürgen R. Reichenbach from University Hospital Jena for insightful discussions. This study is funded by the Independent Research Fund (grant 8020-00158B).

**Abbreviations**

**QSM**: Quantitative Susceptibility Mapping, **SNR:** Signal-to-noise ratio, **FOV**: Field-Of-View, **MEDI**: Morphology-enabled Dipole Inversion, **PDF**: Projection-Onto-Dipoles, **F*l1***: Fitting algorithm with *l1* regularization, **F*l2***: Fitting algorithm with *l2* regularization, **RMSE**: Root-Mean-Squared-Error, **COSMOS**: Calculation of Susceptibility through Multiple Orientation Sampling.

# Figures

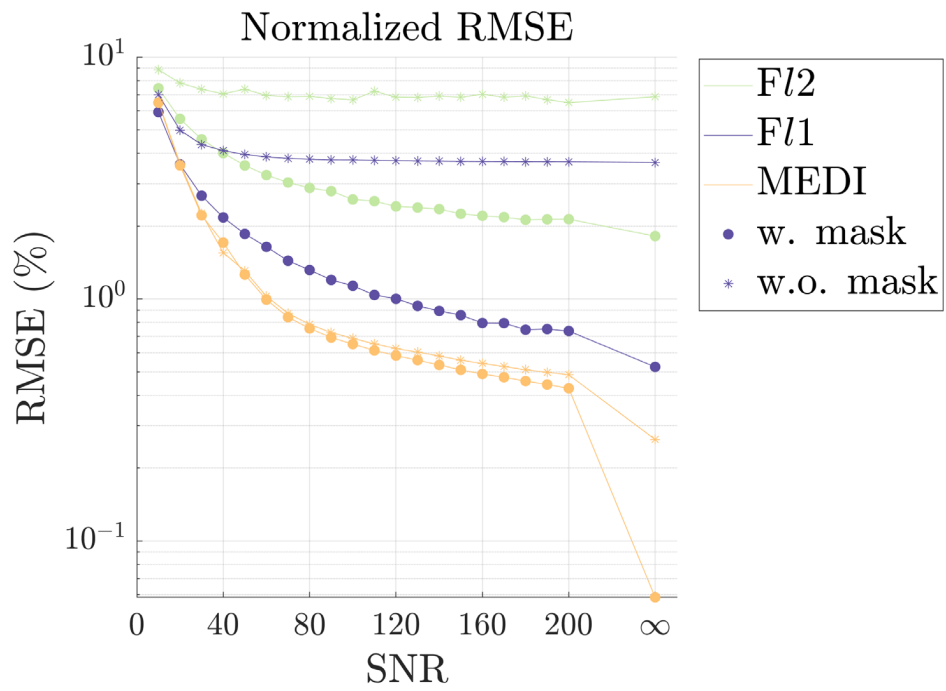

*Figure 1*

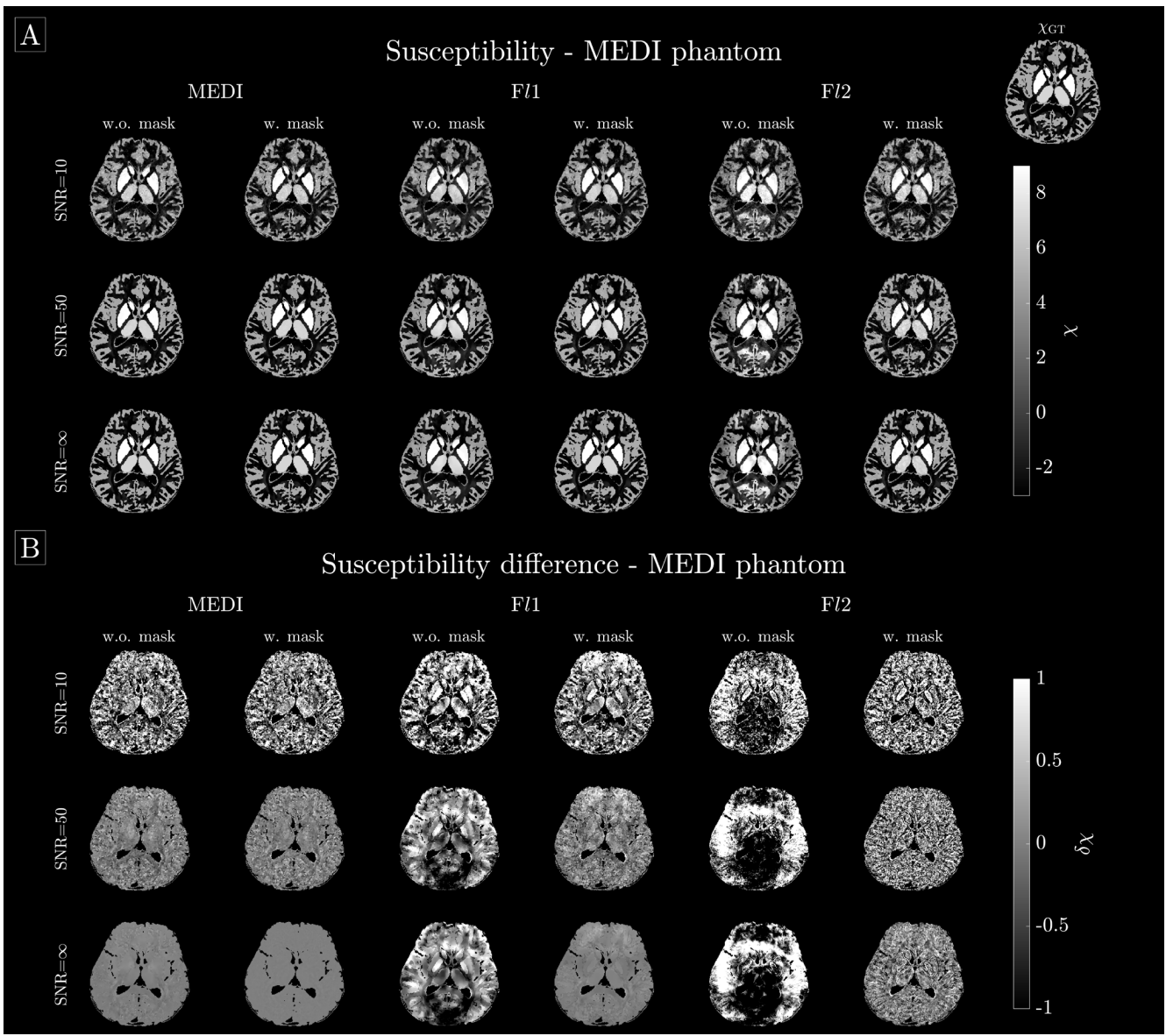

*Figure 2*

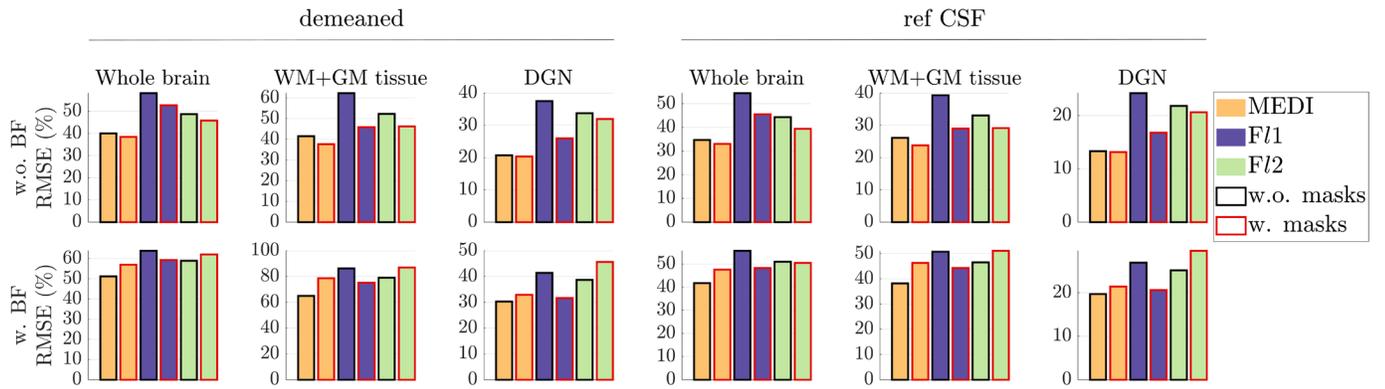

*Figure 3*

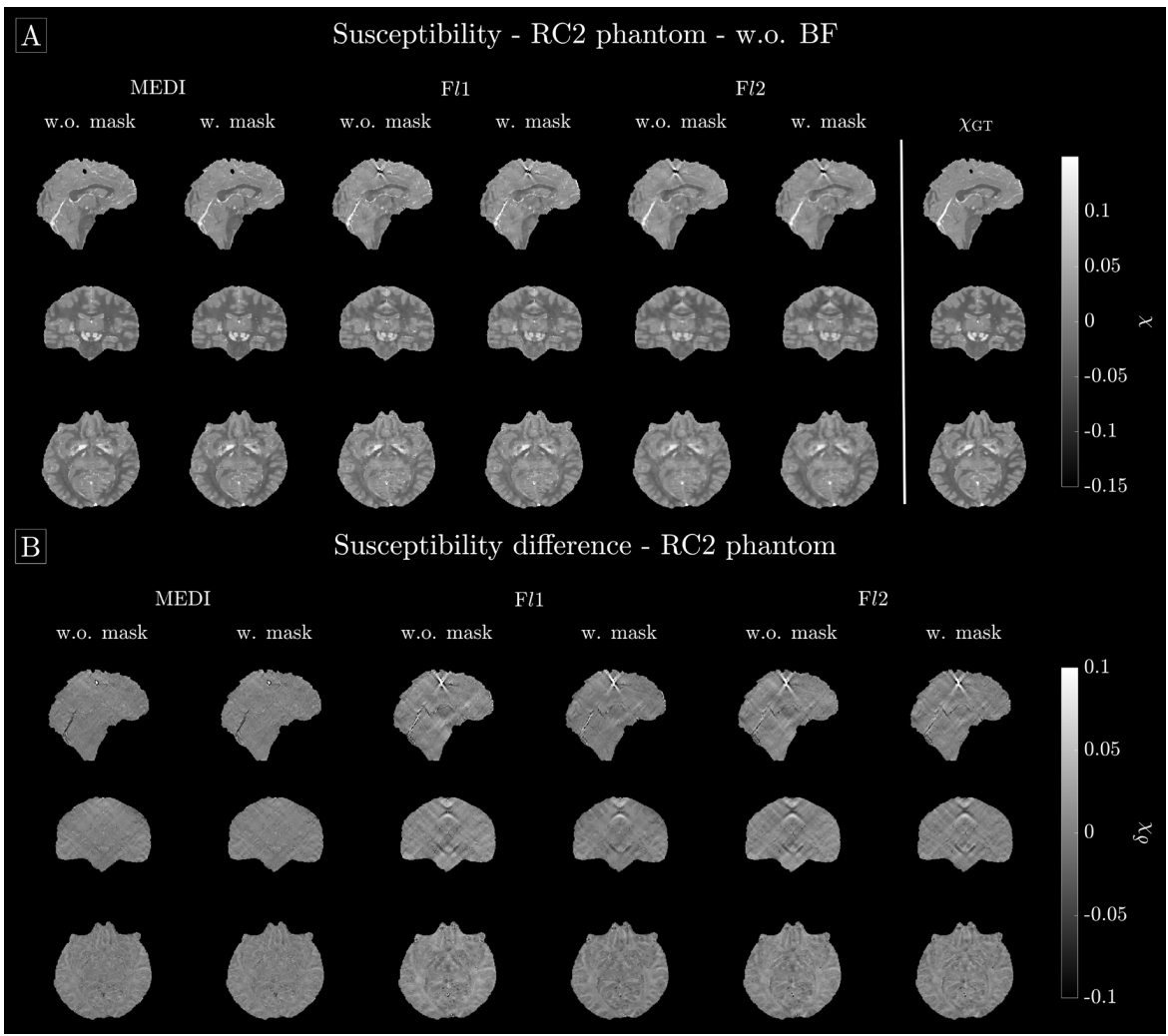

*Figure 4*

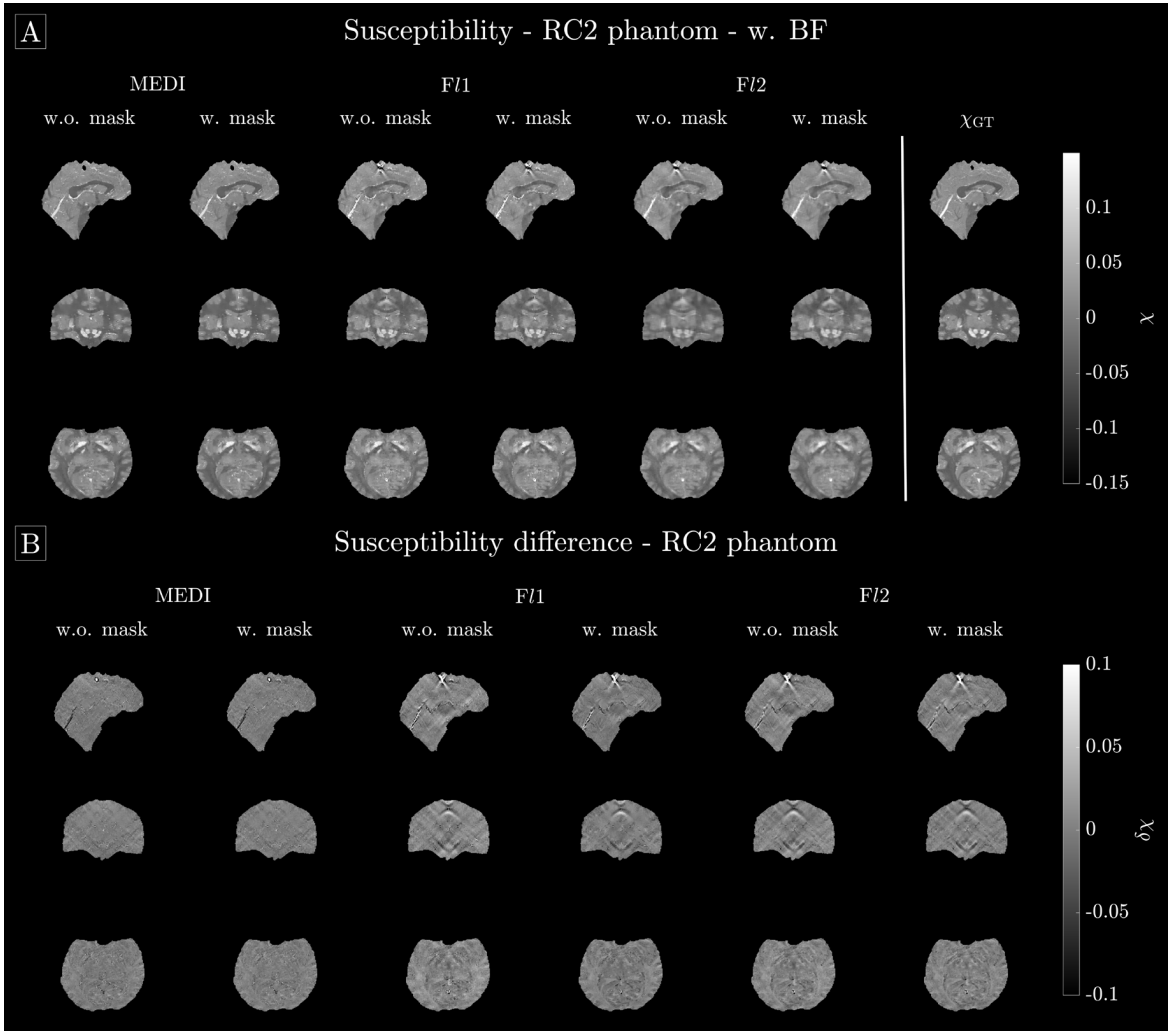

*Figure 5*

# Captions

*Figure 1 - Root-mean-squared error of MEDI phantom:* *The normalized and demeaned RMSE decreases for increasing SNR. The decrease is larger for fitting with masking.*

*Figure 2 - Representative susceptibility maps and their relative errors from a mid-axial slice in MEDI phantom*: ***A*** *shows the resulting susceptibility maps from fitting MEDI phantom with or without masks, using MEDI, Fl1 or Fl2, respectively. Rows corresponds to different SNRs from 10 to ∞.* ***B*** *shows the difference to the ground truth. The color bar is truncated at $\pm 1$ with units in ppm.*

*Figure 3 - Root-mean-squared error of RC2 phantom:* *The normalized RMSE is shown across the whole brain (left), WM and GM (middle) and deep gray matter (right). RMSE is presented either demeaned or referenced to CSF in ventricles. Upper row is RMSE for RC2 phantom without background fields (w.o. BF) and lower is with (w. BF).*

*Figure 4 - Representative susceptibility maps and their relative errors from multiple slice directions in RC2 phantom without background fields (w.o. BF)*: ***A*** *shows the resulting susceptibility maps from fitting RC2 phantom with or without masks, using MEDI, Fl1 or Fl2, respectively.* ***B*** *shows the difference to the ground truth. The color bar is truncated at $\pm 0.1$. Streaking is visible for Fl1 and FL2, but suppressed in MEDI. Otherwise, all maps are visually very similar to ground truth.*

*Figure 5 - Representative susceptibility maps and their relative errors from multiple slice directions in RC2 phantom with background fields (w. BF)*: ***A*** *shows the resulting susceptibility maps from fitting RC2 phantom with or without masks, using MEDI, Fl1 or Fl2, respectively.* ***B*** *shows the difference to the ground truth. The color bar is truncated at $\pm 0.1$. A large erosion of the brain is clearly visible compared to Figure 4. This is due to background field removal, where 8 layers were peeled off. Again, streakings are visible for Fl1 and Fl2, but*

*MEDI also shows misestimation of the calcification (black circle in upper row), which can be seen in the difference map. Besides streaking in Fl1 and Fl2, all maps look visually similar to ground truth.*